\documentclass[%
 sor,
 jor,
 amsmath,amssymb,
 reprint,%
longbibliography
]{revtex4-1}

\usepackage{graphicx}
\usepackage{dcolumn}
\usepackage{bm}

\usepackage{color}
\definecolor {myc} {rgb} {0,0,0}  

\begin{document}

\preprint{AIP/123-QED}

\title[Phase separation dynamics of gluten protein mixtures]{Phase separation dynamics of gluten protein mixtures}

\author{A. Banc\textit{$^{a}$}}
\author{J. Pincemaille\textit{$^{a,b}$}}
\author{S. Costanzo\textit{$^{a}$}}
\author{E. Chauveau\textit{$^{a}$}}
\author{M.S. Appavou\textit{$^{d}$}}
\author{M.H. Morel\textit{$^{b}$}}
\author{P. Menut\textit{$^{b,c}$}}
 \author{L. Ramos\textit{$^{a}$}}
 \email{Laurence.Ramos@umontpellier.fr.}

\affiliation{a. Laboratoire Charles Coulomb (L2C), Univ. Montpellier, CNRS, Montpellier, France}
\affiliation{b. Ing\'{e}nierie des Agro-polym\`{e}res et Technologies Emergentes (IATE), Univ. Montpellier, CIRAD, INRA, Montpellier SupAgro, Montpellier, France}
\affiliation{c. Ing\'{e}nierie Proc\'{e}d\'{e}s Aliments, AgroParisTech, INRA, Universit\'{e} Paris-Saclay, Massy, France}
\affiliation{d. Forschungszentrum J\"{u}lich GmbH, J\"{u}lich Centre for Neutron Science (JCNS) at Heinz Maier-Leibnitz Zentrum (MLZ) Lichtenbergstr. 1, 85748 Garching, Germany.}%

\begin{abstract}

We investigate by time-resolved Synchrotron ultra-small X-ray scattering the dynamics of liquid-liquid phase-separation (LLPS) of gluten protein suspensions following a temperature quench. Samples at a fixed concentration ($237$ mg/ml) but with different protein compositions are investigated. In our experimental conditions, we show that fluid viscoelastic samples depleted in polymeric glutenin phase-separate following a spinodal decomposition process. We quantitatively probe the late stage coarsening that results from a competition between thermodynamics that speeds up the coarsening rate as the quench depth increases, and transport that slows downs the rate. For even deeper quenches, the even higher viscoelasticity of the continuous phase leads to a "quasi" arrested phase separation. Anomalous phase-separation dynamics is by contrast measured for a gel sample rich in glutenin, due to elastic constraints. This work illustrates the role of viscoelasticity in the dynamics of LLPS in protein dispersions.

\end{abstract}

\pacs{XXX}
\keywords{XXX}

\maketitle


\section{Introduction}

Phase-separation phenomena are ubiquitous in condensed matter and play a crucial role in metals, ceramics, semiconductors, complex fluids and biological materials.  Understanding the separation processes is important from
both practical and scientific perspectives. Regarding applications, phase-separation is used to create materials with a bicontinuous morphology that allows a control of the molecular transport as, for instance, in gel permeation chromatography, filtration, catalysis and tissue engineering~\cite{nunes_evidence_1996, ulbricht_advanced_2006,thankamony_preparation_2018}. From a fundamental side, rationalizing phase-separation has generated intensive experimental and theoretical studies since the pioneering work of Cahn and Hillard to describe the process of spinodal decomposition using linearized theory~\cite{cahn_free_1958,cahn_free_1959,cahn_phase_1965}. Dynamic similarity has emerged as a powerful concept: the whole phase-separation dynamics is universal and depends uniquely on one time-dependent length scale and not on the microscopic details of the samples. The validity of this concept is questioned when dynamic symmetry between the two separating phases does not hold anymore. This is typically the case when one of the two phases has a much slower dynamics than the other due to crowding, gelation or vicinity to the glass transition, and is rather common for dispersion of colloids, polymers or proteins in a solvent. Since viscoelasticity plays an important role in the phase-separation dynamics of those complex fluids, the phase separation phenomena occurring in such systems as been referred as viscoelastic phase-separation~\cite{tanaka_unusual_1993, tanaka_viscoelastic_2000}.

In the last decade, phase-separation has been recognized as the principle governing the formation of membraneless organelles in eukaryotic cells. Intrinsically disordered regions in proteins and intrinsically disordered proteins have also been pointed out as playing a role in driving phase transitions in the cell (see the reviews~\cite{shin_liquid_2017, boeynaems_protein_2018} and references therein). However, despite their importance in-vivo, in-vitro studies of the dynamics of phase-separation of intrinsically disordered proteins are still rather scare. Indeed, most studies on dynamics of liquid-liquid phase-separation of proteins mainly concern globular ones~\cite{dhont_spinodal_1996, georgalis_dynamics_1998, tuinier_depletion-induced_2000, tanaka_viscoelastic_2005, cardinaux_interplay_2007, gibaud_closer_2009,da_vela_kinetics_2016, da_vela_arrested_2017}.

Wheat gluten proteins possess intrinsically disordered regions~\cite{rauscher_proline_2006,boire_soft-matter_2019}. Gluten proteins include two classes of polypeptides
differing in their propensity to form intermolecular disulphide bonds. Gliadins, which account for half of the gluten proteins in wheat, are monomeric species and glutenins, the other half,
consist in a concatenation of disulfide bond-stabilized polypeptides whose molecular weight can reach several millions of kDa~\cite{Wrigley}. Gliadins and glutenins share similar amino-acid composition, with high contents in glutamine and proline and a very low content in charged amino-acids~\cite{tatham_wheat_2001}. They all possess unstructured repeated domains rich in glycine and proline, conferring them disorder. Gluten proteins are insoluble in water but soluble in a water/ethanol mixture. In such a solvent, they display both structural and mechanical properties intrinsic to colloids and other properties that are specific of polymers~\cite{van_swieten_size_2003,boire_phase_2013,dahesh_polymeric_2014,dahesh_spontaneous_2016,banc_model_2017}. Therefore, gluten proteins are expected to display a more complex behavior than globular proteins or colloids, since they posses features of proteins, polymers and polymer gels simultaneously~\cite{zhou_why_2018}.

In this paper, we investigate the dynamics of liquid-liquid phase-separation in gluten protein mixtures, in a regime rather concentrated in proteins. Thanks to contrasted protein compositions, we  are able to tune the viscous and viscoelastic properties of the proteins dispersed in a water/ethanol solvent. In this way, we aim at probing the respective role of viscosity and elasticity in the dynamics of phase-separation. We mainly use time-resolved Synchrotron ultra-small X-ray scattering to probe the dynamics of liquid-liquid phase-separation of the samples following a temperature quench. We show that fluid viscoelastic samples depleted in polymeric glutenin phase-separate following a spinodal decomposition process, and that an anomalous phase-separation dynamics is by contrast measured for a gel sample rich in glutenin due to elastic constraints.

\section{Materials and methods}
\subsection{Experimental techniques}
\subsubsection{Rheology~~}

Rheological measurements are performed on a stress-controlled rheometer Physica MCR302 (Anton Paar, Germany). The measuring geometry consists of an upper plate with a diameter of $25$ mm and a lower Peltier plate with a diameter of $50$ mm (H-PTD200) serving as temperature control. The samples are loaded at room temperature. The edges are covered with silicon oil in order to prevent solvent evaporation. After loading, the samples are left to relax and equilibrate at $T=25^\circ$C  for $5$ min before starting the rheological measurements. The protocol used to monitor the evolution of the rheological properties with temperature is as follows: first, a frequency sweep test from $100$ to $0.1$ rad/s is performed at $25^\circ$C in the linear regime (strain amplitude  $8$\%). A  dynamic time sweep test (at frequency $1$ rad/s, strain amplitude $3$\%) is then carried out while the sample is cooled from $25^\circ$C down to $-5^\circ$C, at a cooling rate of $-3^\circ$ C/min. After that, a frequency sweep test is performed at low temperature in the linear regime (strain amplitude  $3$\%). Finally, the samples are heated back up to $25^\circ$C and a frequency sweep test is carried out in the same conditions as the one performed before the temperature ramp, in order to check the reversibility of the phase transition. Strain sweeps tests are also performed on fresh samples at different temperatures in order to assess the linear regime over the whole temperature range spanned in the dynamic thermal ramps.

\subsubsection{Light microscopy~~}

We use an Olympus BX53 microscope equipped with a $40\times$ phase contrast objective (numerical aperture of $0.5$) and a Linkam PE60 stage allowing temperature to be varied between $-20^\circ$C and $90^\circ$C using Peltier elements. The sample is sealed with glue between a microscope slide and a coverslip. The sample thickness is fixed at $50$ $\mu$m  thanks to Mylar spacers. The sample is quenched from $20^\circ$C to the final temperature at a cooling rate of $20^\circ$C/min. Phase contrast images of the sample following the temperature quench are taken using an Olympus DP26 camera every $5$ s during a few hours.

\subsubsection{Small-angle neutron scattering.~~}

Small-angle neutron scattering
(SANS) experiments are performed on KWS2 instrument~\cite{radulescu_studying_2016} operated by the J\"{u}lich Center for Neutron Science at the Heinz Maier-Leibnitz Zentrum (MLZ, Garching Germany) using three
configurations with various wavelengths, $\lambda$, and sample-detector distances, $D$ ($D = 20$ m, $\lambda = 1$ nm, acquisition time $120$ min; $D = 8$ m, $\lambda = 0.7$ nm, acquisition time $20$ min, and $D = 2$ m, $\lambda = 0.7$ nm, acquisition time $2$ min) covering a $q$-range from $0.023$ to $2.9$ $\rm{nm}^{-1}$.  The samples are held in $1$ mm-thick quartz cells. We use a nitrogen flux on the cells to avoid water condensation at low temperature. Standard reduction of raw data is performed by the routine qtiKWS30~\cite{MLZ, banc_small_2016}.

\subsubsection{Ultra-small angle X-ray scattering~~}

Experiments are conducted at the ID02 beamline of ESRF (Grenoble, France). The sample-distance detector is $30$ m, and the wave length $0.0995$ nm, yielding a $q$-range from $1.2\times10^{-3}$ to $6\times10^{-2}$ $\rm{nm}^{-1}$. We use a Frelon (Fast-Readout, Low Noise) detector~\cite{narayanan_multipurpose_2018}. The acquisition time is fixed at $5$ ms. Samples are inserted in sealed quartz capillaries with a diameter of $1$ mm, and placed in a Linkam cell (THMS600/TMS94) allowing temperature to be controlled with a precision of the order of $1^\circ$C. We use a flux of nitrogen gas on the capillary to avoid water condensation at low temperature. The sample structure is probed following a temperature quench from $20^\circ$C to a lower temperature $T_q$ ($T_q$ in the range $14^\circ$C to $-12^\circ$C). The cooling rate is fixed at $-80^\circ$C/min. Hence the time required to reach the final target temperature varies between $4$ s for the highest quench temperature ($T_q=14^\circ$C)  and $28$ s for the lowest one ($T_q=-12^\circ$C).
The dynamics is followed over a duration of about $300$ s with a logarithmic spacing of the data points acquisition so that a large dynamic range can be reached measuring only $\sim 50$ data points.  Between two quenches, the sample temperature is set back to $20^\circ$C. The spectra acquired at room temperature before and after the quench at low temperature are always equal, ensuring that the phase-separation following the temperature quench is reversible, and that no sample damage occurs during measurements. Moreover, repetitive measurements with a $100$ ms exposure time at room temperature on the same position in the capillary gives the same results and ensure that sample damage is not an issue in our experiments. All dynamics tests following a temperature quench are performed on a same capillary, whose position with respect to the incident beam is changed for each temperature quench.
Raw data are analyzed using standard procedures~\cite{sztucki_development_2006}. Absolute scattered intensity (in $\rm{cm}^{-1}$) are obtained by normalizing the data by the sample thickness and by a correction factor determined with a measurement of the scattering of pure water. All data shown in the paper are absolute scattered intensity subtracted by the spectrum at room temperature before the temperature quench, and correspond therefore to excess scattering.

\subsection{Samples}

Protein mixtures are extracted from industrial gluten (courtesy of Tereos
Syral, France). The extraction protocol is adapted from the one previously developed by us to extract protein fractions with different compositions~\cite{boire_phase_2013,dahesh_polymeric_2014}. In brief, $20$ g of gluten powder and $200$ ml of
$50$\% (v/v) ethanol/water are placed in a centrifuge bottle and submitted to a continuous
rotating agitation ($60$ rpm at $20^\circ$C) for $19$ h. After a $30$ min
centrifugation at $15,000$ g at $20^\circ$C, the clear supernatant (protein yield $50$\%) is recovered and placed for $1$ h, in a water bath maintained at a temperature $T_q \leq 12^\circ$C, yielding a liquid-liquid phase-separation between two phases, a pellet enriched in glutenin and a supernatant enriched in gliadin. The pellet and supernatant are immediately
frozen at $-18^\circ$ C before being freeze-dried and ground. The respective volume, concentration and composition of the supernatant and pellet depend on the quench temperature $T_q$~\cite{Thesis-Justine}. We define $Glu$ as the weight fraction of polymeric glutenin in the extract: $Glu=\frac{m_{glu}}{m_{glu}+m_{gli}}$
where $m_{glu}$, resp. $m_{gli}$, is the the mass of glutenin, respectively gliadin, as determined by size exclusion high-performance liquid chromatography. We use here four samples with markedly different protein compositions, $Glu=4$ \%, $44$\%, $57$\% and $66$\%.

Samples are prepared by dispersing the required mass of freeze-dried protein fraction in ethanol/water ($50/50$ v/v). The mixtures are placed in a rotary shaker overnight at room temperature, to ensure full homogenization. Measurements are performed within $6$ days after sample preparation.

\begin{figure}[h]
\centering
  \includegraphics[height=6cm]{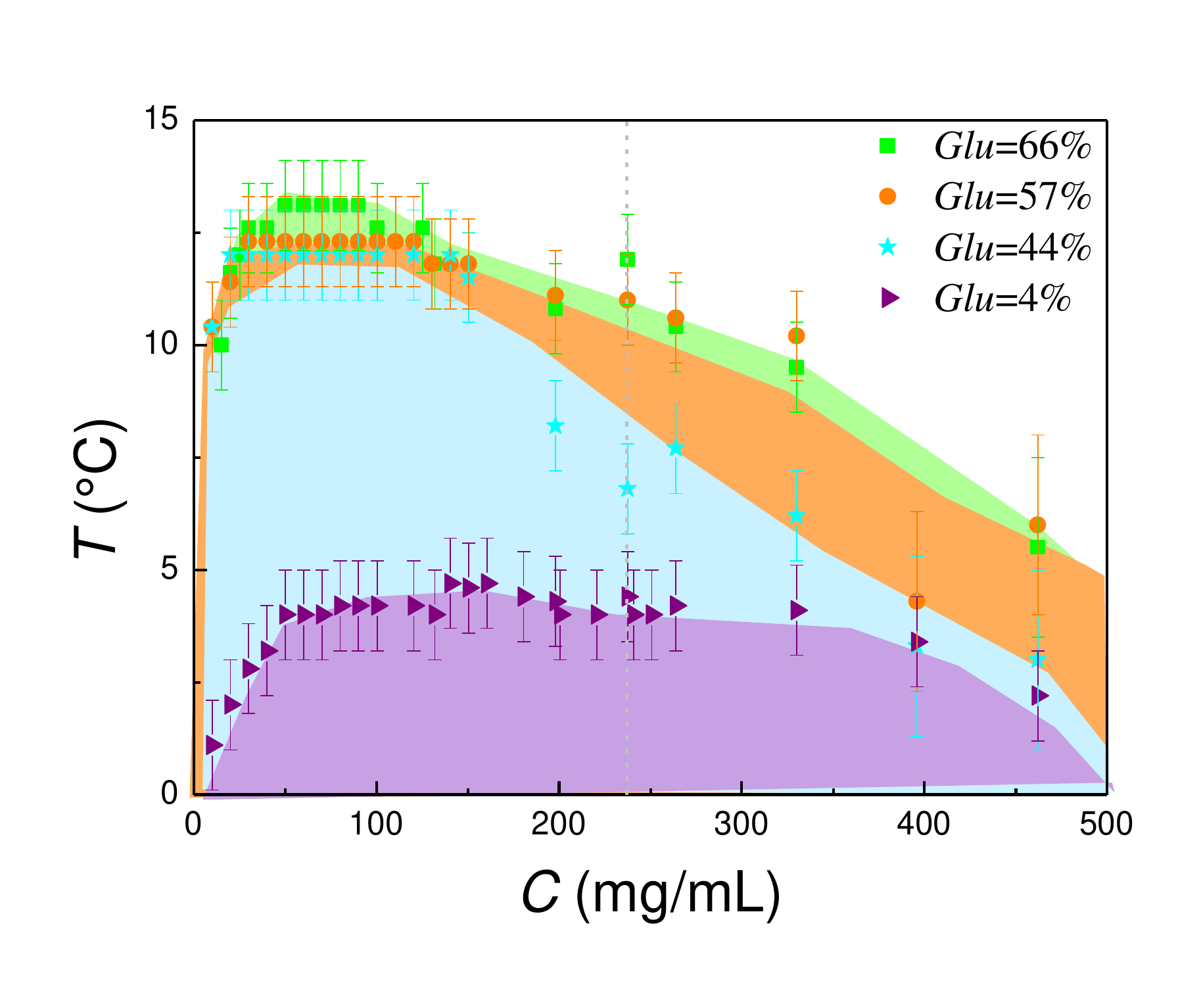}
 \caption{{\color {myc} Phase diagrams for the four protein compositions investigated. The shaded areas show the two-phase regions, and the vertical dashed line shows the sample concentration used to investigate the phase-separation dynamics.}}
 \label{fgr:phasediag}
\end{figure}

{\color {myc} We show in fig.~\ref{fgr:phasediag} the phase-diagrams established through turbidity measurements at low protein concentration, $C$, and differential scanning calorimetry for higher $C$~\cite{pincemaille_methods_2018}, for $C$ in the range $10-500$ mg/ml, and
 for the four protein compositions $Glu=4$ \%, $44$\%, $57$\% and $66$\%}. All samples display an upper critical solution temperature (UCST). We find that the phase-diagrams depend on the protein composition. Note that they are very sensitive to the amount of ethanol in the solvent~\cite{Dill}. Although very difficult to establish precisely especially for the sample rich in gliadin, we find that, for all protein the critical concentration, $C_c$, is of the order of $50-100$ mg/ml. {\color {myc}In the following, we investigate the phase-separation dynamics for a fixed protein concentration $C=237$ mg/ml, which is much higher than $C_c$}, hence the liquid-liquid phase separation yields a protein-rich majority phase and a protein-poor minority phase.

\begin{figure}[h]
\centering
  \includegraphics[height=10cm]{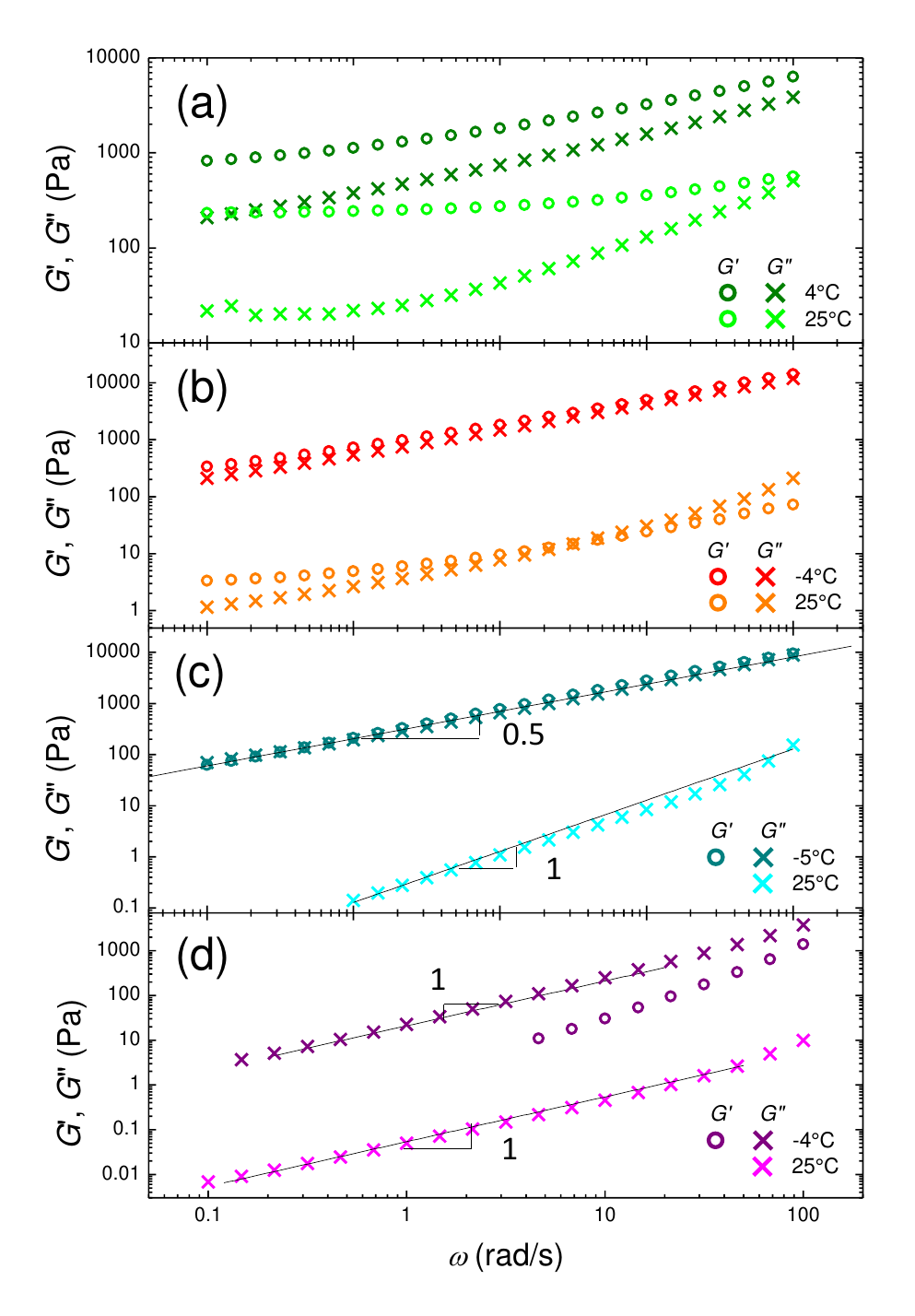}
  \caption{Storage (open circle) and loss (crosses) as a function of the frequency, at temperature above and below phase-separation, for samples with different protein compositions, $Glu=66$\% (a), $57$\% (b), $44$\% (c) and $4$\% (d). Solid lines in (c,d) are power law fits of the experimental data.}
  \label{fgr:rheoLow}
\end{figure}

As well documented in the literature  for wheat flour~\cite{uthayakumaran_basic_2000} and gluten in water~\cite{janssen_rheological_1996,pruska-kedzior_comparison_2008}, sample viscoelasticity strongly depends on the content of glutenin. Our results with model gluten extracts in a mixture of water and ethanol are in line with those results. We show in fig.~\ref{fgr:rheoLow} the frequency-dependence of the storage, $G'$ and loss $G"$ moduli, for all samples at room temperature. The two samples enriched in glutenin are gels, with $G'>G"$ at low frequency, and an elastic plateau $G_0$ which largely depends on the amount of glutenin ($G_0 \simeq 3$ Pa for $Glu=57$\% and $G_0 \simeq 240$ Pa for $Glu=66$\%).  The other two samples, which are depleted in glutenin, are essentially viscous: in the experimentally accessible range of frequency the storage modulus is too low to be measured reliably and the loss modulus is proportional to the frequency yielding viscosities of the order of $1$ Pa.s for $Glu=44$\% and of $50$ mPa.s for $Glu=4$\%.

\section{Experimental results and discussion}
\subsection{Structural and mechanical evidence of phase separation}
\subsubsection{Rheological evidence of phase-separation~~}

The change of the viscoelastic properties upon phase-separation are probed for the different samples by measuring at a fixed frequency ($1$ rad/s) the evolution of the storage and loss moduli as the temperature $T$  decreases from room temperature down to $-5^\circ $C. All samples display qualitatively similar features. Both $G'$ and $G"$ first smoothly increase as $T$ decreases from room temperature, and below a threshold temperature, the moduli increase much more sharply with $T$. However, the way the ratio $G"/G'$ changes with temperature varies from one sample to another. For the purely viscous sample ($Glu=4$\%), viscoelasticity emerges upon temperature decrease as a non-negligible value of the storage modulus is measured below $0\pm1^\circ $C. For the intermediate sample ($Glu=44$\%)
$G'$ increases more than $G"$ and both moduli becomes equal for $T\leq0^\circ$C. For the sample with $Glu=1.3$, $G'$ and $G"$ are roughly equal in the whole temperature range. Finally, we find that $G'$ increases more smoothly than $G"$ as $T$ decreases for the gel sample ($Glu=66$\%) which nevertheless remains mainly elastic in the whole temperature range. We consider the sharp increase of the viscoelastic moduli below a transition temperature $T_t$ ($T_t=6\pm1^\circ $C for $Glu=4$\%, $8\pm1^\circ $C for $Glu=44$\%, $10\pm1^\circ $C for $Glu=57$\% and $12\pm1^\circ $C for $Glu=66$\%) as a signature of the onset of liquid-liquid phase-separation. These values, which might slightly depend on the rate of the temperature ramps, are consistent, within experimental errors, with the phase-diagrams(fig.~\ref{fgr:phasediag}).

The frequency-dependent sample viscoelasticity changes drastically below and above liquid-liquid phase-separation (fig.~\ref{fgr:rheoLow}). Elasticity emerges from the more viscous sample ($Glu=4$\%) which is the more depleted in glutenin although it remains essentially viscous (with a $440$ fold increase of its viscosity, from $50$ mPa.s to $22$ Pa.s). The two samples rich in glutenin, which are gels at room temperature, remain gel-like, with a minor increase of the plateau modulus for the more elastic gel (from $240$ Pa to $830$ Pa for $Glu=66$\%) and an increase by two orders of magnitude, from $3$ to $340$ Pa for the weaker gel with $Glu=57$\%). More remarkably, the intermediate sample ($Glu=44$\%) which is a viscous liquid at room temperature shows at $T=4^\circ $C the typical response of a critical gel~\cite{WinterChambon, martin_viscoelasticity_1988}, with the two moduli following the same power law, $G' \sim G" \sim \omega^{0.5}$.

\begin{figure}[h]
\centering
  \includegraphics[height=6cm]{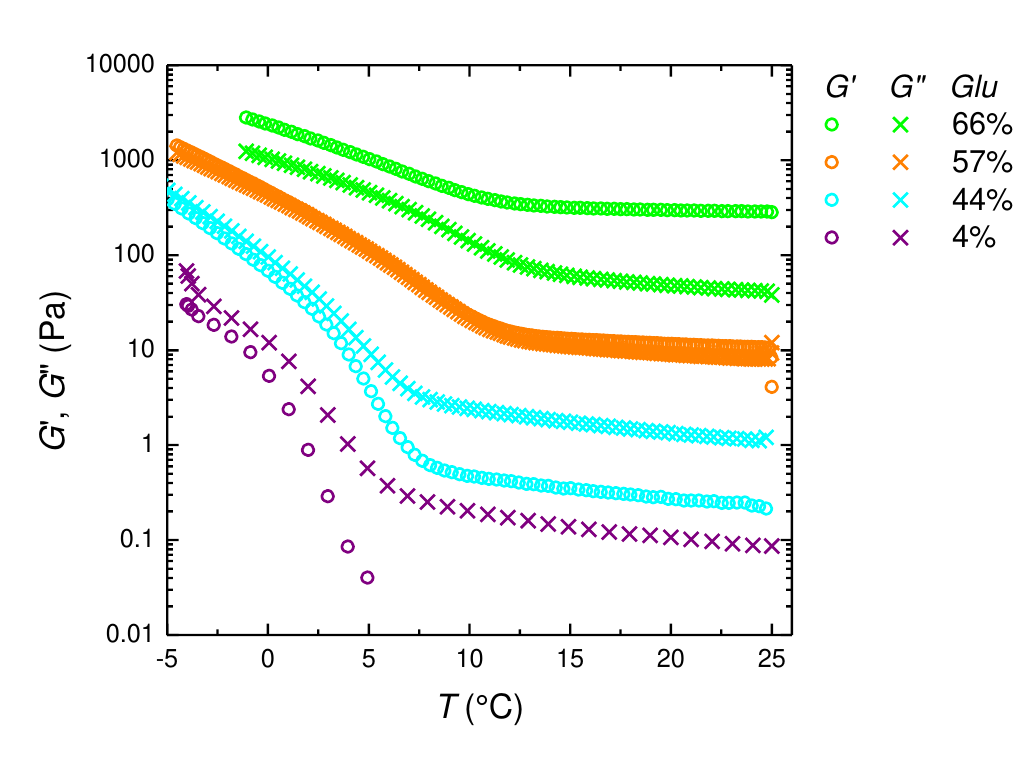}
  \caption{Storage (open circles) and loss (crosses) as a function of temperature for samples with different protein composition, as indicated in the legend. The frequency is fixed at $1$ rad/s and the strain amplitude is $3$ \%. }
  \label{fgr:rheovsT}
\end{figure}

\subsubsection{Imaging of phase-separation~~}

\begin{figure}[h]
\centering
  \includegraphics[height=9cm]{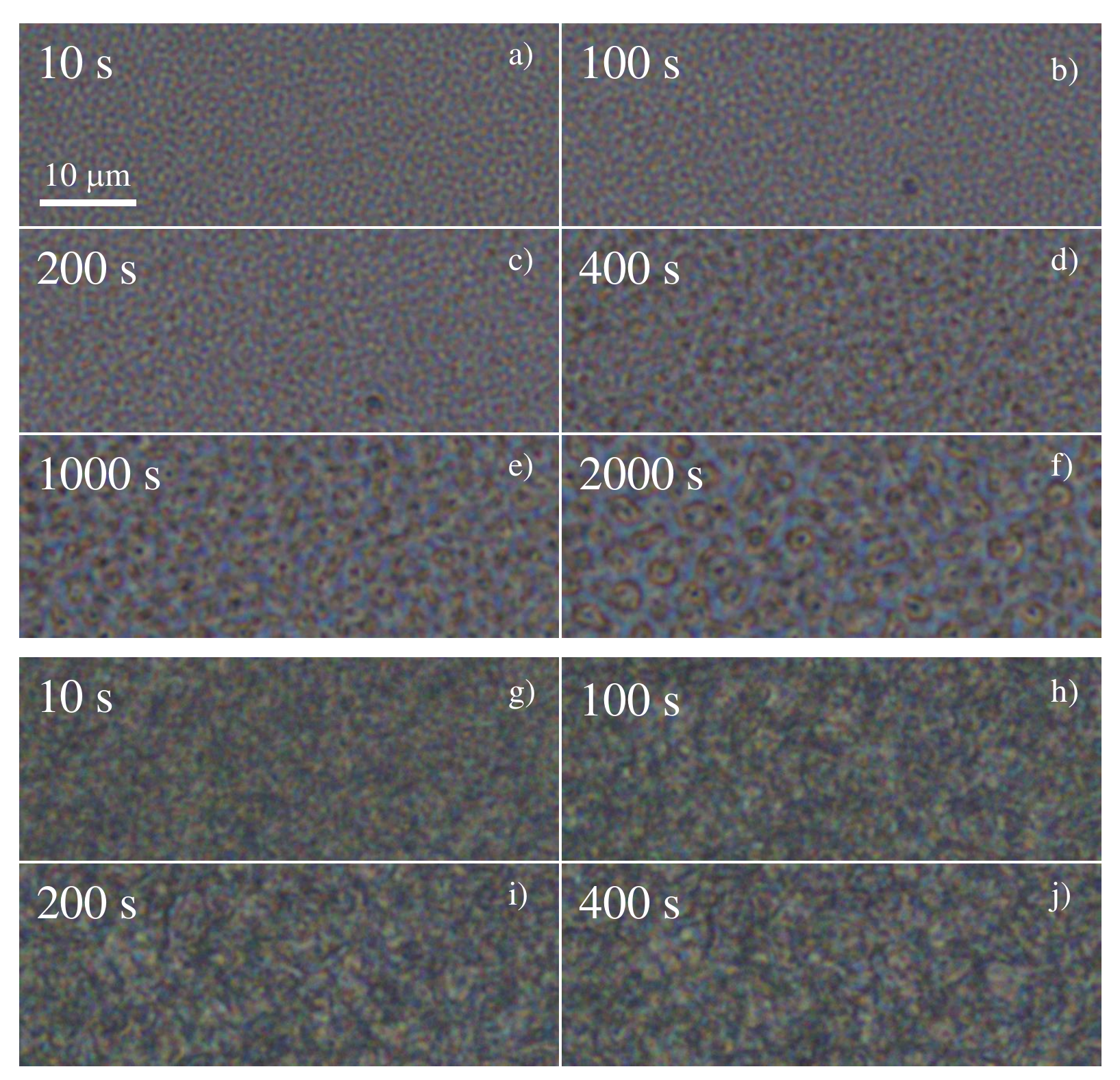}
  \caption{Light microscopy images of a sample with (a-f) $Glu=57$\%, $\Delta T=1^\circ$C, and  (g-j) $Glu=66$\%, $\Delta T=2^\circ$C. The times indicated correspond to the time elapsed since the quench temperature has been reached. The scale is the same for all images.}
  \label{fgr:microscopy}
\end{figure}

Because samples present a UCST, they appear turbid below the transition temperature $T_t$. We note however that, because of the relatively high protein concentration, no macroscopic phase-separation occurs within a few days, and that the turbidity is entirely reversible. Phase-contrast microscopy is used to better visualize the phase-separation processes. We show in fig.~\ref{fgr:microscopy}(a-f) images taken following a temperature quench at a final temperature $T_q=10^\circ$C for a sample with $Glu=57$\%. A bicontinuous morphology, with a characteristic length scale that grows with time, is observed at short time after the temperature quench (fig.~\ref{fgr:microscopy}). Such observation  suggests that the liquid-liquid phase-separation proceeds through a spinodal decomposition process. At long time a percolation-to-droplets transition is observed, as expected for an off-critical mixture~\cite{demyanchuk_percolation--droplets_2004}. Similar features are obtained for the two viscous samples ($Glu=4$ and $44$\%). The pattern observed for the gel sample is more complex (see fig.~\ref{fgr:microscopy}g-j). It does not display a clear regular bicontinuous morphology, as expected for a classical spinodal decomposition, nor droplet as expected for a nucleation and growth process, and does not significantly evolves with time.

\subsubsection{Scattering profiles~~}

\begin{figure}[h]
\centering
  \includegraphics[height=6cm]{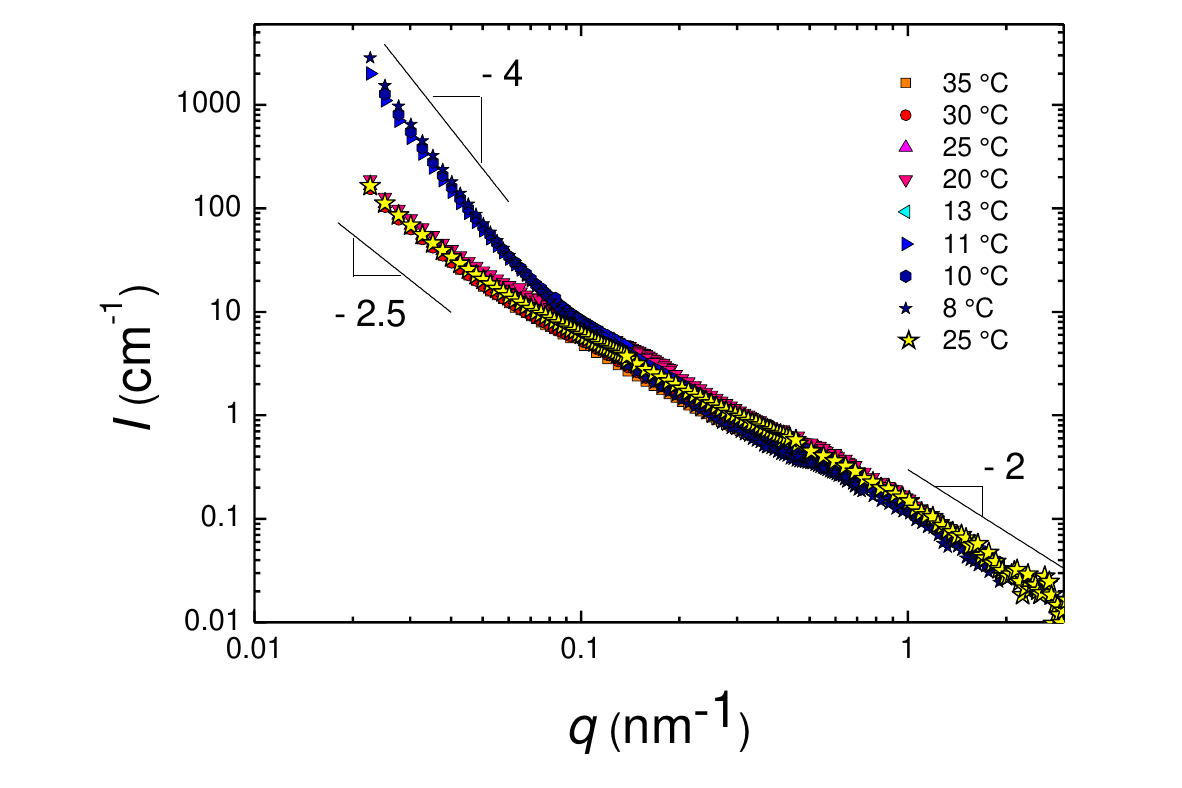}
  \caption{Small-angle neutron scattering patterns measured at different temperatures as indicated in the legend, for a sample with $Glu=66$\%. }
  \label{fgr:SANS}
\end{figure}

We show in fig.~\ref{fgr:SANS}  the small-angle neutron scattering patterns of a sample with $Glu=66$\% at different temperatures from $T=35$ to $8^\circ $C. The scattering patterns for $T$ above the transition temperature $T_t\approx12^\circ$C all perfectly superimpose. They are moreover qualitatively similar to the one already published for $Glu=52$\%~\cite{dahesh_polymeric_2014,banc_small_2016}. At large scattering vectors $q$, the scattered intensity varies as $q^{-2}$, corresponding to the signal of polymer chains in a theta solvent. In addition, the large scale heterogeneities probed at low $q$ for a gel sample translates into a power law scaling $I \sim q^{-p}$, with $p \approx 2.5$. A smooth transition is measured at intermediate $q$ between these two power laws. A striking feature is the sharp modification of the scattered intensity at low $q$, for $T<T_t$, with a transition from a $q^{-2.5}$ scaling to a $q^{-4}$ scaling, which is characteristic of sharp interfaces between the phase-separated protein-rich and protein-poor phases. The transition is measured to occur between $11$ and $13^\circ$C, in full agreement with
{\color {myc} the phase-diagrams (fig.~\ref{fgr:phasediag})} and the rheology data (fig.~\ref{fgr:rheovsT}). Note that the evolution of the scattering pattern is reversible as evidenced by the data acquired at $T=25^\circ$C following the decrease of temperature down to $8^\circ$C, which perfectly superimpose to those acquired before phase-separation. We note that these data have been acquired using small-angle neutron scattering, but we have checked that comparable results are obtained with X-ray scattering (data not shown.)
We mention also that infrared spectroscopy shows that the secondary structures of the proteins is not modified in the phase-separated states: in the one-phase region as in the phase-separated states gluten proteins can be regarded as disordered proteins (data not shown). This is fully consistent with the fact that the local structure of the sample (polymer chains in theta solvent conditions, $I \sim q^{-2}$) is the same above and below transition, as all spectra perfectly superimpose for $q> 0.1$ $\rm{nm}^{-1}$.

\subsection{Dynamics of phase-separation}

As described above, visual observation, light microscopy, rheology and small-angle X-ray and neutron scattering experiments support a liquid-liquid phase-separation, which is reversible, and which does not lead to a macroscopic phase separation of the samples on the time scales of a few days. Quantifying the early stage dynamics of phase-separation following a temperature quench is however difficult with those techniques, as it requires to get time-resolved structural data information on a micrometer length scale. This can be achieved by contrast using time-resolved ultra-small X-ray scattering (USAXS), as described below.

In the following, we define the depth of the temperature quench as $\Delta T= T_t-T_q$ where $T_q$ is the quench temperature and $T_t$ is the transition temperature. By USAXS, for $T>T_t$  the scattering signal does not evolve with time and remains equal to that measured at room temperature, whereas for $T<T_t$, an excess scattering (with respect to the signal measured at room $T$) emerges and increases with time.
Dynamics data are acquired for different depth of temperature quench, for three samples ($Glu=4, 44$ and $66$\%). These three samples allow one to compare the effect of background viscosity for viscous/viscoelastic samples ($Glu=4$\% and $Glu=44$\%) and to assess the role of elasticity when comparing these two samples with the mainly elastic sample ($Glu=66$\%).

We first describe and comment the results for the viscous samples.

\subsubsection{Viscous samples~~}
\paragraph{Spinodal decomposition and coarsening.~~}

\begin{figure}[h]
\centering
  \includegraphics[height=6 cm]{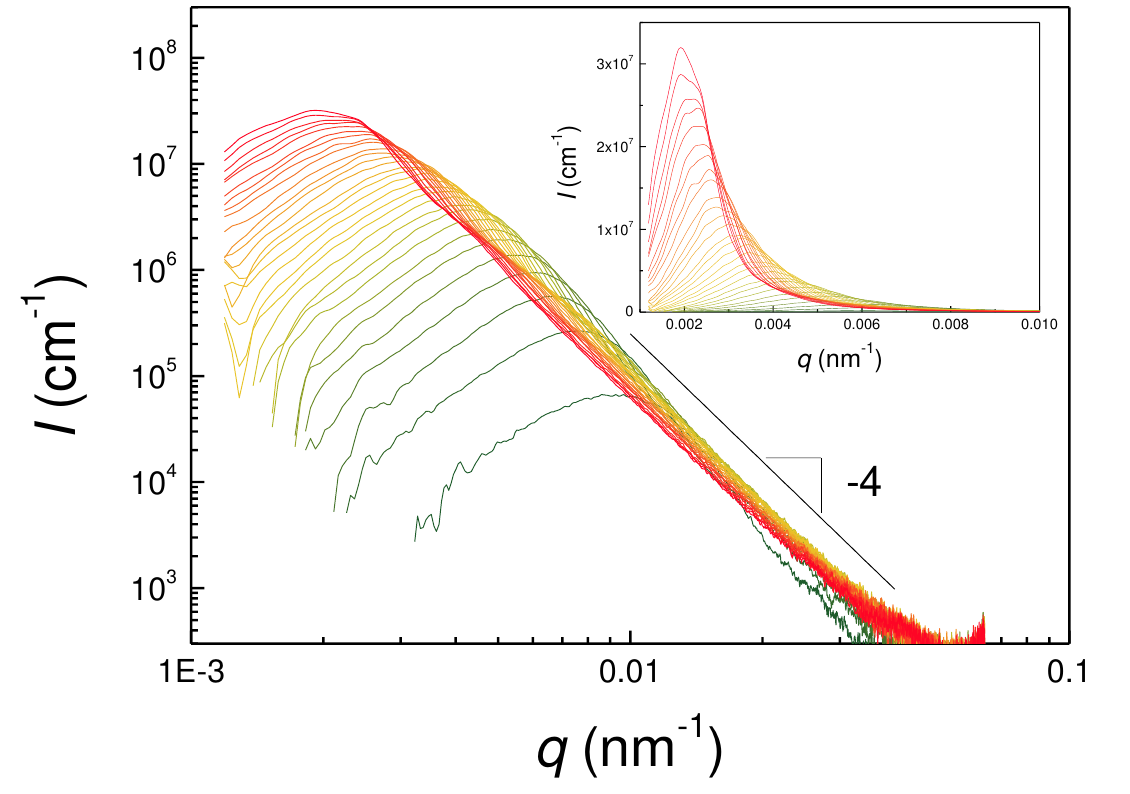}
  \caption{Time evolution of the scattered intensity for a sample with $Glu=44$\% following a temperature quench of depth $\Delta T=7^\circ$C. The time elapsed since the quenched temperature has been reached varies from $1$ to $300$ s, when the color gradually changes from green to red.}
  \label{fgr:I08}
\end{figure}

We start describing in details the results obtained for the sample with $Glu=44$\%. As an illustration, we show in fig.~\ref{fgr:I08} the time evolution of the excess scattering patterns following a temperature quench with $\Delta T=7^\circ$C, but we mention that similar results are obtained for $\Delta T$ in the range $(1-15)^\circ$C. By construction, no signal is measured before phase-separation, since the scattered intensity shown is the excess of scattered intensity with respect to the scattered intensity of the sample at room temperature. Very rapidly after the target temperature has been reached, or even before the final temperature is reached for the lowest temperature investigated, a clear excess scattered intensity is measured, and a well defined peak emerges. Concomitantly, a $q^{-4}$ scaling is measured at large $q$ that is the signature of the existence of sharp interfaces between two phases: $I=Bq^{-4}$.  With time, we find that the peak position $q_{\rm{max}}$ shifts towards lower wave vector, and that the height of the peak, $I_{\rm{max}}$ {\color {myc} increases}, whereas the Porod prefactor $B$  varies non-monotonically.

\begin{figure}[h]
\centering
  \includegraphics[height=6cm]{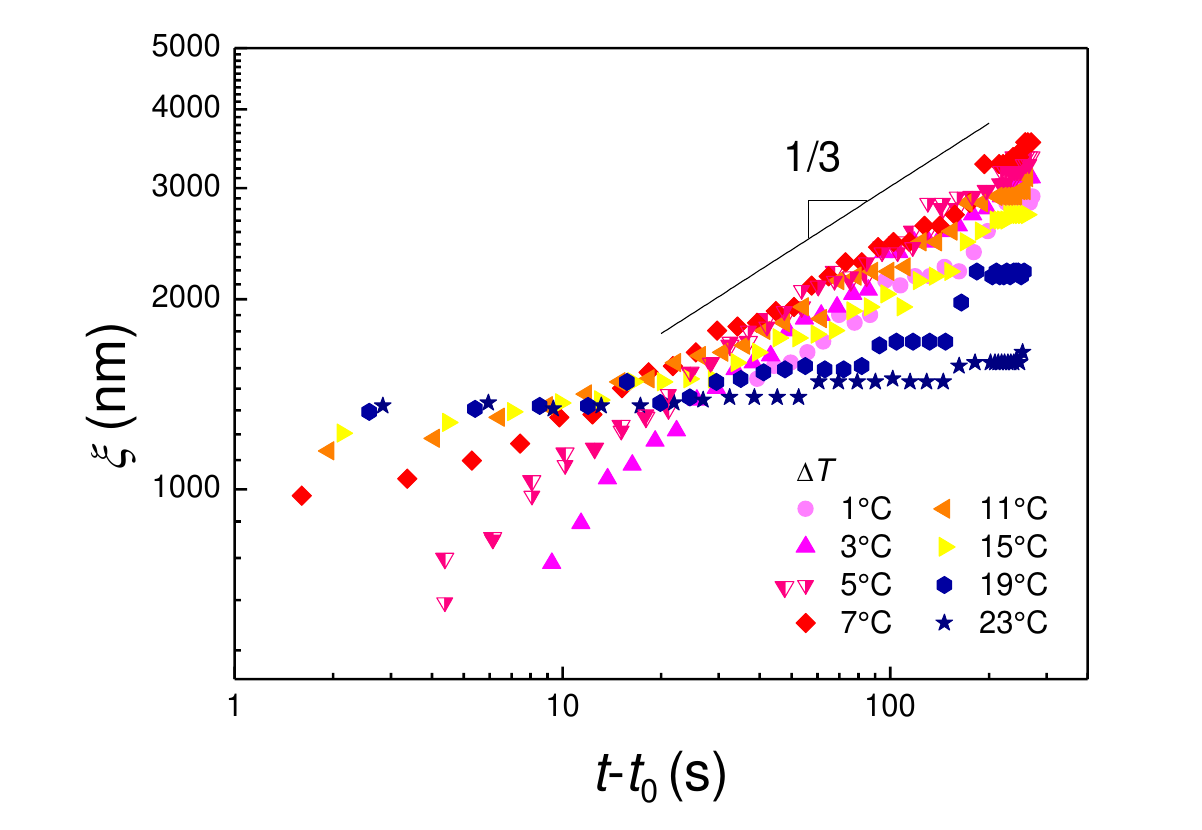}
  \caption{Characteristic length $\xi$ as a function of the time elapsed since the quench temperature has been attained, for a sample with $Glu=44$\%. Different symbols correspond to different quench depths, as indicated in the legend.}
  \label{fgr:Xi08}
\end{figure}

The existence of a peak indicates a preferential length scale in the phase-separated sample. This length scale can be defined as $\xi=2\pi/q_{\rm{max}}$. We show in fig.~\ref{fgr:Xi08} the time evolution of $\xi$ for different quench depths $\Delta T$. Data are plotted as a function of $t-t_0$, the time elapsed since the temperature has reached its target value ($t$ is the time at which the quench starts and $t_0$ is the time at which the target temperature is reached). We also show on the same plot two data sets of the time evolution of $\xi$ for a same temperature quench ($\Delta T=5^\circ$C). The second set of measurements has been acquired after the sample have been submitted to several quenches (with $\Delta T=7, 15, 19, $ and $23^\circ$C) and hence have undergone several phase-separations. The fact that the two sets of data perfectly superimpose over the whole time scale shows that the phase-separation is reversible on the range of temperature and time scales investigated here and does not lead to macroscopic phase separation, in agreement with visual inspection and light microscopy. For $\Delta T$ between $1$ and $15^\circ$C, we find that at long time (i.e. for $t-t_0 \geq 10$ s), $\xi$ varies as a power law with time: $\xi=A(t-t_0)^{m}$, with a same exponent $m=1/3$ for all $\Delta T$. This dynamics is the one expected for the late stage coarsening of a spinodal decomposition  in the case of diffusive growth (when hydrodynamic flow does not play any role) by either coarsening or Oswald ripening~\cite{Lifshitz, binder_theory_1974, Siggia}.

\begin{figure}[h]
\centering
  \includegraphics[height=7cm]{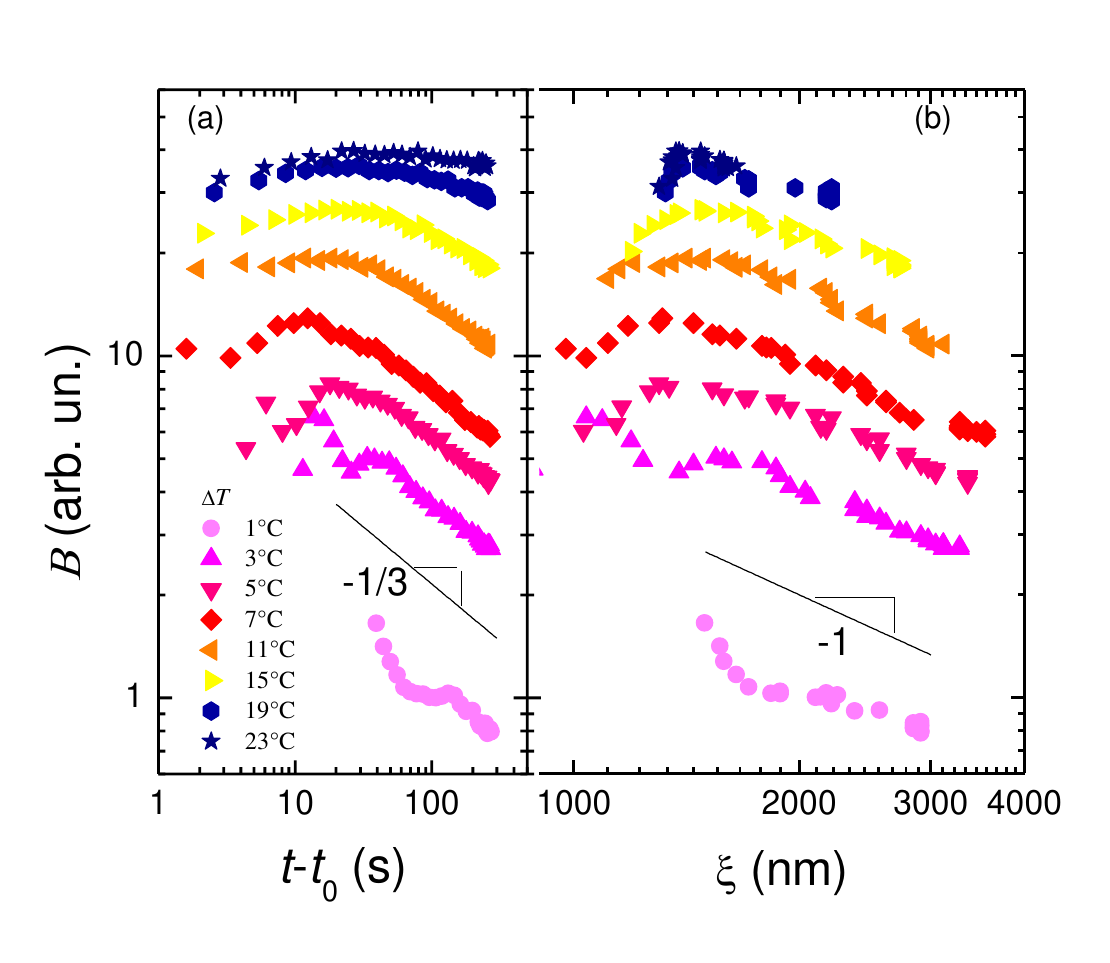}
  \caption{Porod prefactor as a function of (a) time and (b) the characteristic length, for a sample with $Glu=44$\% quenched at different depths
  as indicated in the legend.}
  \label{fgr:Porod08}
\end{figure}

On the other hand, the Porod prefactor $B$ reads: $B\sim (\Delta \rho)^2 \frac{S}{V}$, where $S/V$ is the specific surface of the interface (with $S$ the total surface area between the two phases and $V$ the sample volume) and $\Delta \rho$ is the contrast between the two phases, which mainly depends on the respective protein concentration between the two phases. Figure~\ref{fgr:Porod08} shows the evolution of $B$ for phase-separations following quenches of different depths, $\Delta T$. We measure that $B$ increases with $\Delta T$ due to increasing contrast between the two phases as the protein-rich phases is expected to become even richer and the protein-poor phase even poorer when the quench is deeper. With time a non-monotonic evolution of $B$ is measured. At longer time, we find that $B$ varies as a power law with time with an exponent $-1/3$ for $\Delta T \leq 15^\circ$C (fig.~\ref{fgr:Porod08}a). In the late stage coarsening, when the two phases have reached their equilibrium concentration and the contrast between the two phases $\Delta \rho$ is constant, we expect $S \sim N \xi^2$ where $N$ is the number of patterns of size $\xi$ with $N=V/\xi^3$. Hence, $S/V \sim B \sim 1/\xi$ as measured experimentally (fig.~\ref{fgr:Porod08}b).  By contrast, we attribute the increase of the Porod prefactor at short time to an intermediate stage of spinodal decomposition, where the characteristic length $\xi$ varies very smoothly with time while the contrast between the two phases increases, hence leading to an increase of $B$.

\begin{figure}[h]
\centering
  \includegraphics[height=12cm]{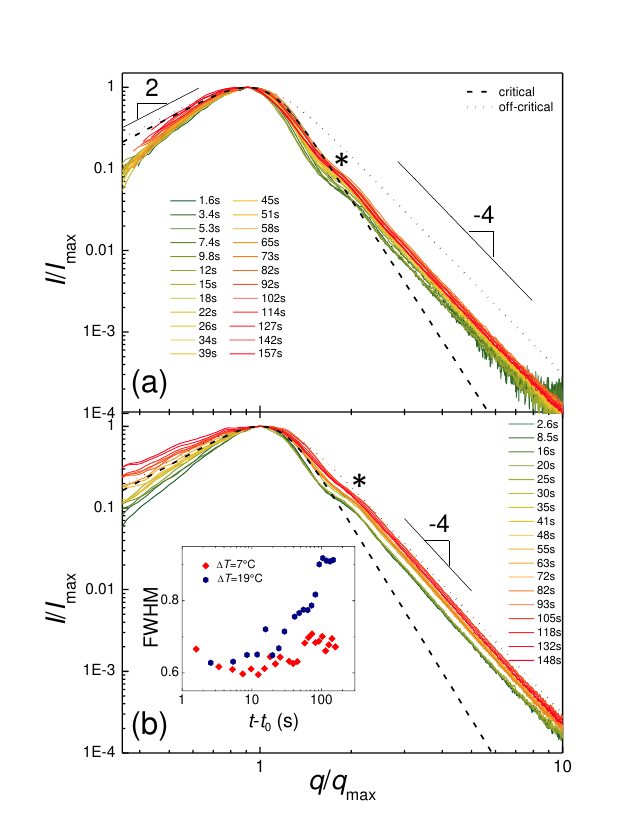}
  \caption{Time evolution of the scattered intensity plotted in normalized units, for a sample with $Glu=44$\% with a temperature quench depth of (a) $7^\circ$C, (b) $19^\circ$C. In (a,b) the experimental data are the colored lines and the Furukawa prediction for critical and off-critical mixtures (see text) are the thin dashed and dotted black lines. The asterisks point to the position of the shoulder. (b, inset) Time evolution of the full width at half maximum of the data shown in the main two plots.}
  \label{fgr:DynSim}
\end{figure}

Dynamic similarity, which implies that only one time-dependent length determines the evolution of morphology, is a hallmark of spinodal decomposition. We show below that it holds for our experiments. Indeed, by plotting the scattering curve in rescaled units, where the scattered intensity is normalized by the peak value, $I_{\rm{max}}$, and the wave vector by the peak position, $q_{\rm{max}}$, we find a nice superposition over the whole time window of the $I/I_{\rm{max}}$ vs $q/q_{\rm{max}}$ plots, as shown in fig.~\ref{fgr:DynSim}a for a quench depth $\Delta T=7^\circ$C. Furukawa has proposed a simple empirical law to account for the shape of the peak: $I/I_{\rm{max}}=\frac{[1+\frac{\gamma}{2}]x^2}{\frac{\gamma}{2}+x^{2+\gamma}}$ with $x=q/q_{\rm{max}}$, and $\gamma=6$ for a critical mixture, and  $\gamma=4$ for an off-critical mixture~\cite{furukawa_dynamics-scaling_1984}.  We find that these laws reproduce reasonably well our experimental data.  For $x<1$, we find the $q^2$ predicted by Furukawa for both critical and off-critical mixtures. In addition, for $x>2$, we find the $q^{-4}$ Porod scaling proposed in the case of off-critical mixtures. Moreover, at all times, we observe that the scattering data exhibit a less intense second peak, a shoulder, located at about $2q_{\rm{max}}$. Such shoulder has been observed in the intermediate or late stage coarsening of liquid and polymer binary (critical or off-critical) mixtures~\cite{Bates, cumming_nucleation_1990,kubota_spinodal_1992,kuwahara_spinodal_1992,kuwahara_kinetics_1993}.

\paragraph{Towards arrested phase-separation.~~}

\begin{figure}[h]
\centering
  \includegraphics[height=6cm]{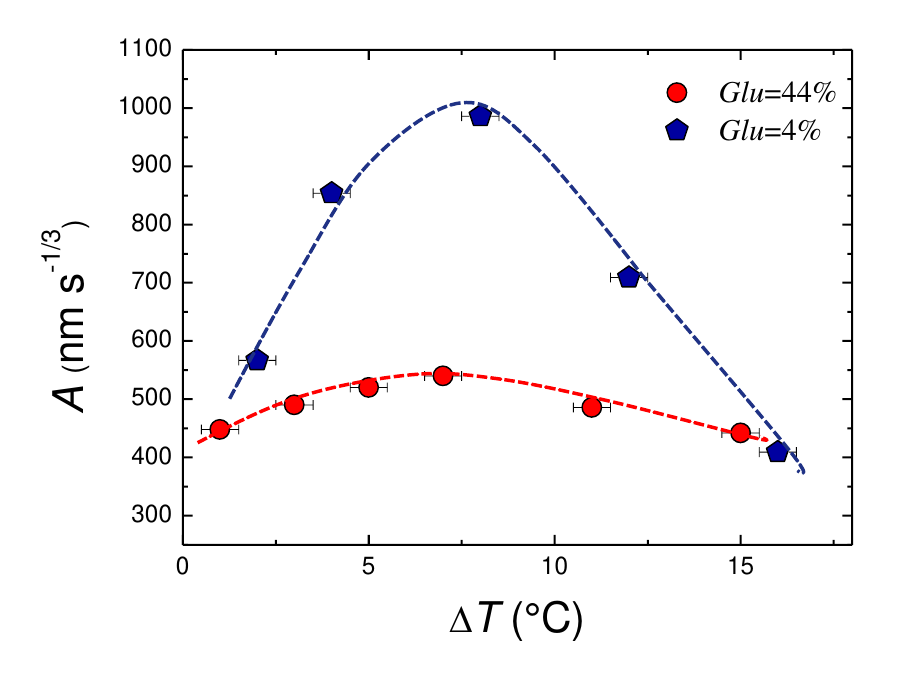}
  \caption{Growth rate of the characteristic length as a function of quench depth, for two samples with $Glu=4$\% and $Glu=44$\%. The dashed lines are guides for the eye.}
  \label{fgr:Rate}
\end{figure}

In contrast to the evolution of the Porod prefactor with quench depth $\Delta T$, we interestingly find that the prefactor characterizing the coarsening rate, $A$, varies non-monotonically with $\Delta T$ (fig.~\ref{fgr:Rate}). As observed and theoretically predicted for a polymer solution near a glass transition~\cite{barton_dynamics_1998}, the non-monotonic behavior is the signature of a competition between thermodynamics, which tends to speed up the phase-separation hence increase the rate as $\Delta T$ increases, and transport, which tends to decrease the rate when $\Delta T$ increases, due to the larger viscosity of the majority phase. When the quench becomes even deeper, the protein concentration of the continuous protein-rich phase may become so high that this phase become highly viscous or even elastic, impeding further coarsening. Accordingly, for very large $\Delta T$, we find that $\xi$ almost does not vary over the whole duration of the experiment, suggesting an arrested phase-separation (fig.~\ref{fgr:Xi08}). Interestingly, however, we find that the scattering profiles still evolve significantly during the stage where the peak position very weakly changes. This is shown using the rescaled units in fig.~\ref{fgr:DynSim}b for a quench depth $\Delta T=19^\circ$C. Here a log-log plot of $I/I_{\rm{max}}$ vs $q/q_{\rm{max}}$ clearly evidences, as time evolves, an evolution of the scattering profiles at low $q$, i.e. for $q/q_{\rm{max}}<1$. With time, the peak becomes less and less marked as the scattering intensity at low $q$ continuously increases. This can be quantified by measuring the full width at half maximum (FWHM) of the peak in the rescaled units. We find that FWHM continuously increases with time at $\Delta T=19^\circ$C whereas it is roughly constant for $\Delta T=7^\circ$C (inset fig.~\ref{fgr:DynSim}b). This indicates that the so-called arrested state, as inferred from the time evolution of the peak position,  is not fully arrested and that, with time, the size distribution of the phase-separated domains becomes increasingly wider.
We mention that the same evolution of the scattering patterns has been observed for polymer melts~\cite{ochi_phase_2015}.

We finally note that qualitatively similar results are obtained for the less viscous sample ($Glu=4$\%) (data not shown). Interestingly, we also observe a non-monotonic evolution of the growth rate of the characteristic length $\xi$ with the quench depth (fig.~\ref{fgr:Rate}). Overall however, the rates are larger for the less viscous sample ($Glu=4$\%), as expected.

\subsubsection{Gel sample~~}

\begin{figure}[h]
\centering
  \includegraphics[height=6cm]{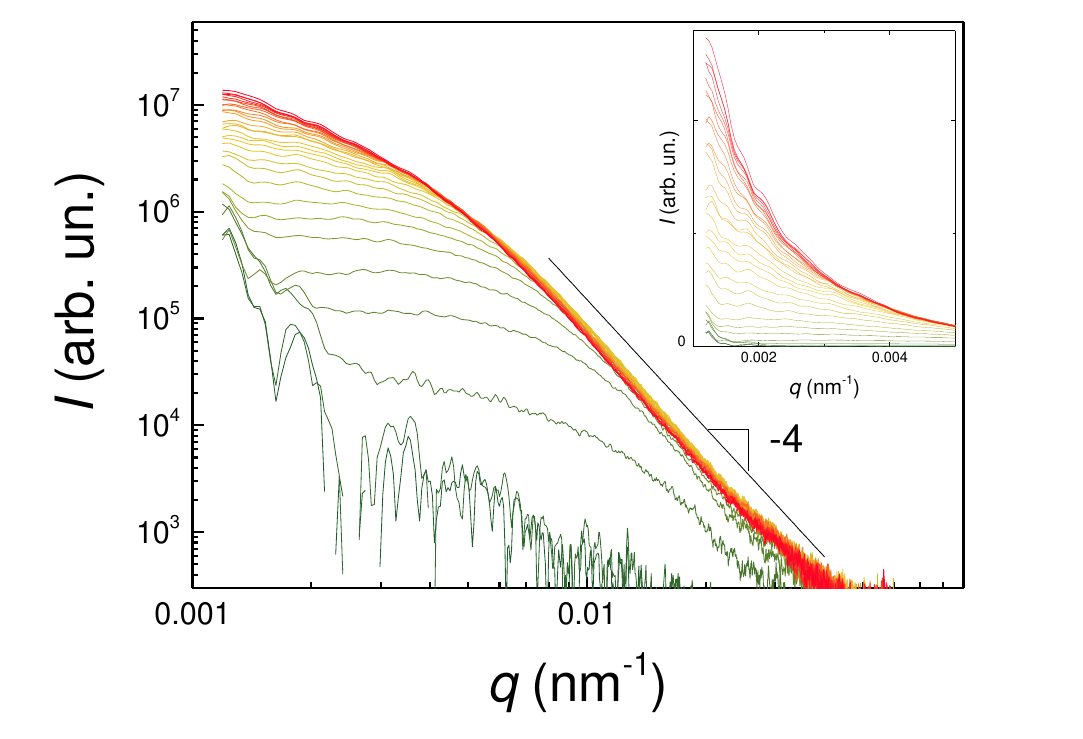}
  \caption{Time evolution of the scattered intensity, in a log-log plot (main graph) and in a lin-lin plot (inset) for a sample with $Glu=66$\%, following a quench of depth $1^\circ$ C. The time elapsed since the quenched temperature has been reached varies from $10$ to $300$ s, when the color gradually changes from green to red.}
  \label{fgr:Igel}
\end{figure}

The dynamics of the phase-separation of the gel exhibits distinct features from those of the viscous samples. The time evolution of the excess scattering pattern following a temperature quench is shown in fig.~\ref{fgr:Igel}. In line with the data obtained for viscous fluids, a $q^{-4}$ scaling emerges very rapidly after the temperature quench and signs the occurrence of sharp interfaces between two phases. However, in marked contrast with previous samples, no peak in the scattering pattern is observed, at any time. Instead a plateau-like evolution of the excess scattering is measured at small $q$.

\begin{figure}[h]
\centering
  \includegraphics[height=7cm]{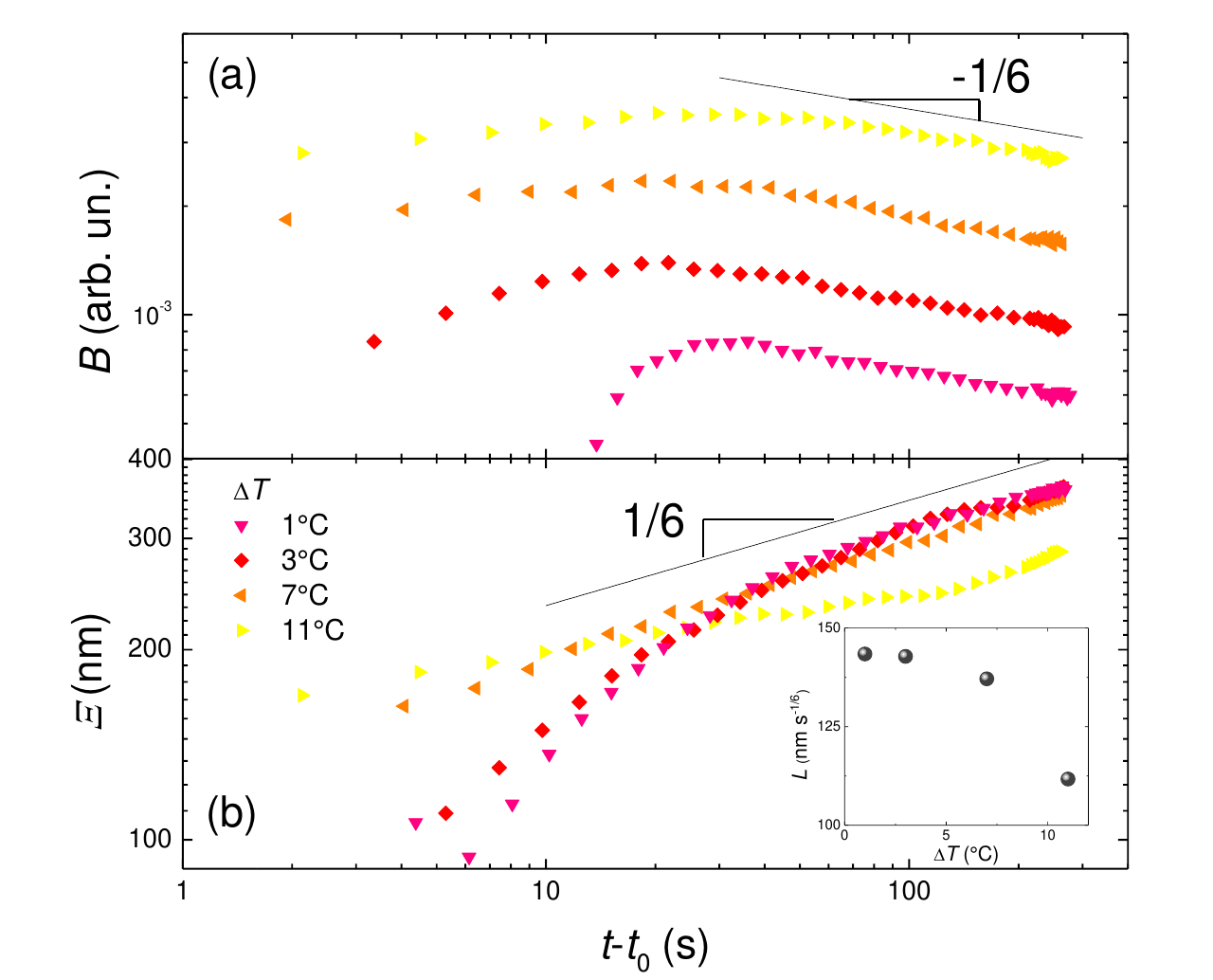}
  \caption{Porod prefactor (a) and characteristic size (b) as a function of time for a sample with $Glu=66$\% quenched at different temperatures as indicated in the legend. Inset: prefactor of the growth rate of the characteristic size as a function of the quench depth.}
  \label{fgr:Porod19}
\end{figure}

We first analyze the amplitude of the Porod scaling factor, $B$, as a function of time for various quench depths $\Delta T$ (fig.~\ref{fgr:Porod19}a). Overall $B$ increases with the quench depth, as expected from a larger protein concentration contrast when $\Delta T$ increases, but the same features are measured for all $\Delta T$ for the time evolution of $B$. The Porod prefactor displays a non-monotonic evolution with time: At short times, $B$ increases with time (the increase being smoother for deeper quench), and at later times, by contrast, $B$ decreases with time, signing the late stage coarsening process. In this regime, a same scaling, $B \sim (t-t_0)^{-m}$ with $m=0.17 \pm 0.1 \approx 1/6$, is measured for all $\Delta T$. Hence, the time evolution measured here is slower than the one expected for late stage coarsening of spinodal decomposition as measured by us for viscous samples (with $m=1/3$).

On the other hand, we find that the excess scattering profiles are well fitted by a Debye-Bueche (DB) model: $I=(B \Xi^4)/[1+(q \Xi)^2]^2$. Such model has been introduced to describe micro-phase-separated solids with sharp interfaces~\cite{DebyeBueche} and is conventionally applied to inhomogeneous media in general, in particular to account for static inhomogeneities in polymer gels (see the review~\cite{seiffert_scattering_2017} and references therein). Figure~\ref{fgr:Porod19}b shows the time evolution of the characteristic length scale $\Xi$ for different quench depths $\Delta T$. In the late stage coarsening, we measure $\Xi=L(t-t_0)^{-m}$, with $m=1/6$ (fig.~\ref{fgr:Porod19}b) and with a prefactor $L$ that monotonically decrease as $\Delta T$ increases (inset of fig.~\ref{fgr:Porod19}b).
As observed for viscous fluids, we find also here consistent time evolution of the Porod prefactor and the characteristic length scale $\Xi \sim 1/B$, with however a power law exponent $m=1/6$ much smaller than the one expected for growth controlled by diffusion as measured for fluid samples. Similarly, we attribute the increase with time of $B$ in the early stage to an increase with time of the contrast between the two phases.
We mention that the $1/6$ power law has been observed for binary polymer mixtures~\cite{tanaka_unusual_1993, kuwahara_kinetics_1993} and has been interpreted as due to viscoelastic phase-separation~\cite{tanaka_viscoelastic_2000}, which here would be associated to the slow dynamics of glutenin polymers. The $1/6$ exponent is also consistent with theoretical predictions for spinodal decomposition of solids~\cite{binder_theory_1974,furukawa_dynamics-scaling_1984}. However, we stress here that we do not observe a peak, signing a well defined characteristic length scale, in the coarsening process. The length $\Xi$ that is extracted from a DB model corresponds to a cut-off length in a length scale distribution.

At first sight, the scattering pattern we observe (a continuous decrease of the scattering intensity with the wave-vector) might be attributed to a nucleation and growth process. In Refs.~\cite{hashimoto_time-resolved_1984,nunes_evidence_1996,graham_dynamics_1997,lefebvre_fluctuation_2002} the absence of peak was mentioned for shallow quenches in polymer/solvent mixtures. However, no quantitative analysis of the scattered intensity has been performed preventing any quantitative comparison with the Debye-Bueche data analysis described here. In addition, in our case, the same phenomenon is measured for all quench depths, whereas a nucleation and growth process is expected to take place only for shallow quenches. Moreover, nucleation and growth is usually a slow process, whereas here an excess scattering is measured as soon as the quench temperature is reached. Furthermore, although difficult to quantitatively analyze, the light microscopy images of the phase-separation reveals a homogeneous yet complex pattern, but not isolated droplets (fig.~\ref{fgr:microscopy}).
For all these reasons, we believe that the features observed during the phase-separation of the gluten gel do not correspond to a process akin to a nucleation and growth phenomenon, but instead could correspond to an anomalous spinodal decomposition due to the solid-like nature of the continuous phase.

To the best of our knowledge, there are few works that investigate phase-separation dynamics in a soft elastic solid. Our work is very different from Ref.~\cite{style_liquid-liquid_2018} which considers the liquid-liquid phase-separation of the solvent swelling a polymer network. In our case, this is the protein that forms the viscoelastic network which tend to phase-separate. Intuitively, one therefore expects that the elastic constrains due to the network prevent a standard liquid-liquid phase-separation through spinodal decomposition, although an apparently standard spinodal decomposition has been observed in a chemically cross-linked polymer gel~\cite{bansil_kinetics_1996}. Our observations are by contrast in line with turbidity and ultrasonic measurements that suggest for chemically cross-linked gel large size distribution of heterogeneities due to phase-separation~\cite{hu_spinodal_2001}. This has to be connected to theoretical arguments regarding the fact that cross-links acts as pinning point which likely prevent global ordering~\cite{massunaga_phenomenological_1997}. Clearly, further work would be desirable to better characterize and model phase-separation processes in an elastically constrained environment.

\section{Conclusions}

We have investigated the dynamics of liquid-liquid phase-separation of gluten protein mixtures using mainly ultra-small angle X-ray scattering techniques.  This technique, which allows one to gather quantitative data with a fast acquisition rate on the micrometer length scale without multiple scattering issues (as opposed to light scattering), has been complemented with light microscopy and rheology measurements to reach a global understanding of the liquid-liquid phase-separation processes that take place in viscous, viscoelastic and gel protein mixtures.

We have probed the dynamics of phase-separation of protein systems following a temperature quench. We have used off-critical mixtures, with a protein concentration larger than the critical concentration. This implies that one crosses the metastable zone, where a nucleation and growth process takes place, before reaching the unstable zone at $T_q$.  However, because the kinetics of nucleation and growth processes are slow and the experimental quenches are relatively fast, we do not expect any perturbation of the scattering signal by these processes, as argued in~\cite{lefebvre_determination_2002,cabral_spinodal_2018}. Accordingly, we have not evidenced any signature of nucleation and growth process even for the shallowest quench investigated, as opposed to what has been observed for gliadin suspensions by light scattering~\cite{boire_dynamics_2018}. The phase-separation dynamics of viscous protein suspensions present all the hallmarks of classical spinodal decomposition. We could not properly characterize the early stage of spinodal decomposition, but instead we have nicely followed the late stage coarsening. The characteristic size extracted from the spinodal ring follows a $1/3$ power law with time, an evolution theoretically expected for a diffusion-controlled growth process. For the first time for this class of proteins, we have evidenced the effect of the overall viscoelasticity of the suspensions in the phase-separation dynamics.
We have shown that the growth rate dependence on the quench depth follows a non-monotonic dependence that results from a balance between thermodynamic and transport processes. This competition will ultimately lead to an arrested phase-separation as the quench depth increases, likely due to the increased viscosity of the continuous phases, as observed for globular proteins~\cite{cardinaux_interplay_2007, gibaud_closer_2009,da_vela_kinetics_2016,da_vela_arrested_2017}, polypeptides~\cite{glassman_arrested_2015}, polymers~\cite{barton_dynamics_1998} and colloids~\cite{lu_gelation_2008,zhang_phase_2013, sabin_exploring_2016, tsurusawa_formation_2017}. Interplay between kinetics of phase separation and kinetics of gelation has been investigated in particular in biopolymer mixtures~\cite{tromp_kinetics_1995, owen_rheology_1998}. We believe however that gelation of the continuous phase is probably not an issue in our case as we have shown for similar wheat gluten protein mixtures that gelation is a slow process~\cite{dahesh_spontaneous_2016}.
On the other hand, we have evidenced an anomalous liquid-liquid phase-separation in the gel samples, which is characterized by a slow coarsening dynamics and a very broad size distribution of the phase-separating patterns. {\color {myc} Our results strongly suggest that } those features result from the elastic constrains provided by the gel structure of the proteins.

Interestingly, for both viscous and gel samples, we have presented a quantitative analysis of the Porod prefactor of the scattering patterns, which account for the total amount of sharp interfaces in the sample. Such analysis provides a nice consistency of the results extracted from the spinodal ring for viscous samples, and for the cut-off length scale for gel samples. Hence, although not commonly performed to the best of our knowledge, a quantification of the Porod prefactor provides an interesting alternative, in particular when the $q$-range of the scattering data does not allow one to access the characteristic length scale.

In addition to the viscoelastic behavior, one remarkable feature of our samples is that they are multi-component systems. To the first order, they are composed of a mixture of monomeric gliadins and polymeric glutenins. Phase separation in multicomponent systems is expected to be more complex to that occurring in binary mixtures. Quite unexpectedly, our experimental results suggest that the overall dynamics of phase-separation does not seem to be strongly perturbed as one recovers, for viscous samples, what has been found for simple suspensions of globular proteins. Nevertheless, we know that upon phase separation the concentrated phase gets enriched in glutenins whereas the diluted phases gets enriched in gliadins. We have leveraged on this complexity~\cite{Thesis-Justine} to produce samples with markedly different compositions and in particular to obtain protein extracts enriched in glutenin, the polymeric proteins which are responsible for the unique viscoelastic properties of wheat dough. In this optics, texturing samples based on wheat protein thanks to arrested liquid-liquid phase-separation, as done with other food products~\cite{donald_physics_1994,mezzenga_understanding_2005, bhat_spinodal_2006, gibaud_new_2012}, could provide a versatile way to texture food products based on gluten proteins.

\section*{Conflicts of interest}
There are no conflicts to declare.

\section*{Acknowledgements}
We acknowledge the European Synchrotron Radiation Facility (ESRF), Grenoble, France, for
provision of synchrotron radiation facilities. We would like to thank Theyencheri Narayanan and Alessandro Mariani for assistance in using beamline ID02 at ESRF. This work is also based upon experiments performed at the KWS-2 instrument operated by JCNS at the Heinz Maier-Leibnitz Zentrum (MLZ), Garching, Germany. Theyencheri Narayanan is acknowledged for his critical reading of the manuscript. The Labex Numev (ANR-10-LAB-20) is acknowledged for financial support.






%
\end{document}